\documentclass{aa}   

\usepackage{subfig}
\usepackage{graphicx}
\usepackage{txfonts}
\usepackage{natbib}
\begin{document}

   \title{Anharmonicity in the mid-infrared spectra of polycyclic aromatic hydrocarbons:}

   \subtitle{Molecular beam spectroscopy and calculations}

   \author{A. K. Lemmens
          \inst{1,}
          \inst{2}
          \and
          D. B. Rap
          \inst{2}
          \and
          J. M. M. Thunnissen
          \inst{2}
          \and
          C. J. Mackie
          \inst{3}
        \and
          A. Candian
          \inst{4}
          \and
          \newline A. G. G. M. Tielens
          \inst{4}
          \and
          A. M. Rijs
          \inst{2}
          \and
          W. J. Buma
          \inst{1,}
          \inst{2}
          }

   \institute{Van't Hoff Institute for Molecular Sciences, University of Amsterdam, Science Park 904, \newline 1098 XH Amsterdam, The Netherlands\\
              \email{w.j.buma@uva.nl}
         \and
            Radboud University, Institute for Molecules and Materials, FELIX Laboratory, \newline Toernooiveld 7c, 6525 ED Nijmegen, The Netherlands\\
             \email{e-mail: a.rijs@science.ru.nl}
              \and
            Chemical Sciences Division, Lawrence Berkeley National Laboratory, Berkeley, California 94720, USA\
            \and Leiden Observatory, Leiden University, Niels Bohrweg 2, 2333 CA Leiden, The Netherlands \\}

\date{Received date /
Accepted date }

% \abstract{}{}{}{}{} 
% 5 {} token are mandatory
 
  \abstract
   {}
   {In this work we determine the effects of anharmonicity on the mid-infrared spectra of the linear polycyclic aromatic hydrocarbons (PAHs) naphthalene, anthracene, tetracene and pentacene recorded using the free electron laser FELIX. }
   {Comparison of experimental spectra obtained under supersonic jet conditions with theoretically predicted spectra was used to show if anharmonicity explicitly needs to be taken into account. }
   {Anharmonic spectra obtained using second-order vibrational perturbation theory agree on average within 0.5\% of the experimental frequencies. Importantly, they confirm the presence of combination bands with appreciable intensity in the 5-6  $\mu$m region. These combination bands contain a significant fraction of the IR absorption, which scales linearly with the size of the PAH. Detection and assignment of the combination bands are a preliminary indication of the accuracy of far-IR modes in our anharmonic theoretical spectra. Detailed analysis of the periphery-sensitive CH out-of-plane band of naphthalene reveals that there is still room for improvement of the VPT2 approach. In addition, the implications of our findings for the analysis of the aromatic infrared bands are discussed. }
{}


   \titlerunning{Anharmonicity in the Mid-IR Spectra of PAHs}
\authorrunning{A.K. Lemmens et al.}
\maketitle
%
%________________________________________________________________
\section{Introduction}

  Currently, the generally accepted hypothesis for the origin of the aromatic infrared bands (AIBs) is the IR emission of UV-pumped polycyclic aromatic hydrocarbons (PAHs) \citep{Allamandola1989,Sellgren1984,Tielens2008}. Because of their ubiquity, PAHs can have major effects on heating and ionization processes in the interstellar medium (ISM), and thereby on the chemistry occurring in molecular clouds \citep{Herbst2001}. Vice versa, the spectral features of these compounds are in principle sensitive probes for the physical environment of particular regions in the ISM and their chemical evolution. 
The relative strength of the observed AIBs implies that astronomically relevant PAHs have sizes of the order of 50 carbon atoms \citep{Andrews2015,Boersma2010,Croiset2016,Ricca2010}. Our knowledge on the photophysical properties of such large PAHs is as yet far from complete, but is key to advancing the interpretation of astronomical observations \citep{Tielens2013}. Laboratory spectroscopy is therefore needed to gain more insight into these properties \citep{Oomens2011}. Such experiments should preferably be carried out under astronomically relevant conditions, that is, under low-temperature and isolated conditions such that spectral shifts and/or intensity variations by the environment can be excluded \citep{Joblin1994}. 

In recent years we have pioneered an approach in which we employ mass-selective ion detection in combination with high-resolution laser spectroscopy to obtain mass- and conformation-selective IR absorption spectra with a resolution and sensitivity that by far exceed what was hitherto possible. This work has primarily focused on the 3 $\mu$m near-IR region associated with aromatic and aliphatic CH stretch vibrations, and has shown how extensive the influence of anharmonicity \citep{Candian2017} is on the appearance of the spectra \citep{Huneycutt2004,Maltseva2016}. State-of-the-art anharmonic calculations were shown to be able to accurately reproduce the experimental spectra \citep{Mackie2015a,MacKie2016,Mackie2018}. This conclusion is important as it provided strong and unambiguous support for the use of theoretical spectra calculated with such a methodology in cases where experimental spectra are not available. 

The studies on the near-IR region were used to determine the presence of specific classes of PAHs based on the overall shape of the near-IR emission. A more detailed characterization -- and potentially a probe for identifying individual PAHs -- is in principle available by considering the mid- and far-infrared spectral features \citep{Boersma2010}. The mid-IR part of the spectrum of neutral gas-phase PAHs has so far, however, remained largely unexplored, with the exception of individual emission bands of naphthalene \citep{Pirali2009}. Mid-IR studies have been performed employing matrix isolation spectroscopy (MIS)\citep{Hudgins1998}, but these spectra are subject to spectral shifts, intensity distortions \citep{Maltseva2018} and sample impurities \citep{Joblin1994}. A further aspect impeding gas-phase mid-IR studies is the availability of appropriate light sources. Here, we resort to molecular beams and use a Free Electron Laser to obtain cold, gas-phase IR absorption spectra of isolated molecules that can be used without further modification to validate theory. 
The far-IR (below $\pm$ 660 cm$^{-1}$) is another spectral regime containing signatures that can be directly related to the identity of specific PAHs \citep{Boersma2010,Ricca2010,Zhang1996}. The interpretation of the spectral features in this region heavily relies on properly taking anharmonic effects into account and validation of calculated far-IR frequencies of PAHs is therefore necessary. Anharmonic bands in the mid-IR can be associated with combination bands built upon one or more far-IR fundamentals. Mid-infrared spectra may therefore aid in the validation of both mid-IR and far-IR calculated vibrational frequencies. 

In the present study we obtained the mid-IR spectra of the full range of  polyacenes from naphthalene to pentacene. This series gives us the possibility to assess the effect of molecular size on photophysical properties and to extrapolate these results to larger species. It will be shown that anharmonic calculations lead to theoretically predicted spectra that are in good (0.5\% deviation) agreement with the experimental ones. Combined, these experimental and theoretical studies underpin the importance of anharmonicity in this region of the spectrum, and provide key handles for furthering the development of astronomical models.

%__________________________________________________________________

\section{Methods}

Experiments have been performed at the FELIX laboratory \citep{Oepts1995} using the molecular beam setup described in \citep{Rijs2011}. Samples of naphthalene, anthracene, and tetracene were heated close to their melting point in a glass container and expanded into vacuum with a Series 9 pulsed valve from General Valve using argon as a carrier gas. The backing pressure was typically around 2.5 bar. The low vapor pressure of pentacene at our maximum heating temperature required the use of laser desorption \citep{Meijer1990,Rijs2015} to get sufficient molecules into the gas-phase. The sample was mixed with graphite powder and applied on a solid graphite bar. A pulsed 1064 nm laser beam (Polaris Pulsed Nd:YAG Laser, NewWave Research) with a typical energy of 1 mJ/pulse was used to desorb the pentacene molecules. Subsequently they were picked up by a jet of argon (4 bar backing pressure) created with a Jordan pulsed valve. The skimmed molecular beam was crossed by a UV excitation laser beam with typical pulse energies of a few millijoules that was provided by a Nd:YAG laser pumped UV dye laser (Spectra-Physics/Radiant Dye), an ionization laser beam of 193 nm provided by an ArF excimer laser (Neweks) and an IR laser beam provided by the free electron laser (FEL) FELIX. The ions were mass-separated using a time-of-flight spectrometer and detected with a multichannel plate detector. A delay generator (SRS DG645) was used for synchronization and triggering of the equipment. Infrared spectra were acquired by IR-UV ion dip spectroscopy, tuning the UV excitation laser beam to the S$_{1}$ $\leftarrow$ S$_{0}$ transition \citep{Amirav1980,Maltseva2016} of the investigated compound and scanning the IR wavelength. The bandwidth of the IR source FELIX is about 1\% of the photon frequency produced. 

Experimental spectra have been compared to simulated spectra on the anharmonic \citep{Barone2005,Barone2014} B3LYP \citep{Becke1993}/N07D \citep{Barone2008,Mackie2018} level of theory -- hereafter referred to as anharmonic -- and the harmonic B3LYP \citep{Becke1993}/4-31G level of theory -- hereafter referred to as harmonic. The anharmonic spectra were calculated using the Gaussian16 suite of programs using default thresholds \citep{Frisch2016}, whereas the harmonic spectra were downloaded from the NASA Ames PAH database and scaled with multiple scaling factors as described in \citet{Bauschlicher2018}. For comparison with the experimental spectra, theoretical spectra have been convolved with a Gaussian line shape with a full width at half maximum (FWHM) of 0.3\% of the photon frequency (slightly narrower than the FEL bandwidth of 1\% for clarity). 

\section{Results}

Figures 1a-d report the mid-IR spectra of jet-cooled naphthalene, anthracene, tetracene, and pentacene, respectively, in the 5-18 $\mu$m region. The linewidths are determined by the bandwidth of FELIX, which is approximately 1\% of the photon frequency. This means that a narrower linewidth is obtained in the far-IR part of the spectrum (about 3 cm$^{-1}$ FWHM) than in the 5 $\mu$m region. The experimental spectra are compared to both harmonically calculated spectra obtained from the NASA Ames PAH spectral database \citep{Bauschlicher2018, Ricca2010} (Figures 1a-d in blue) that are conventionally  used, and spectra obtained from high-level anharmonic calculations (Figures 1a-d in green, present work). The spectral features with a large absorption match reasonably well with the spectra predicted using the harmonic approximation. The strongest band by far in the experimental spectrum of naphthalene (Figure 1a) is at 783 cm$^{-1}$ and can be assigned to a CH out-of-plane (OOP) bending mode involving all CH bonds. The experimental spectra of longer polyacenes (anthracene, tetracene and pentacene, Figures 1b-d) display in this frequency region two intense bands with comparable intensities. The lower-energy band corresponds to the CH OOP bending mode of the four adjacent hydrogens on the outer aromatic rings (quartets), while the band on the blue side corresponds to the CH OOP bending mode of the solo hydrogens on the inner aromatic rings. Since naphthalene does not possess enclosed rings, it displays only one single band. 

The spectra show that the intensity of the solo CH OOP modes increases upon elongation of the polyacene chain. This observation confirms that the CH OOP frequency region is sensitive to the periphery of the molecule considered \citep{Bellamy1975}. The CH OOP bands of pentacene are relatively broad, causing the absorption intensity to be spread over a larger frequency range. Although the integrated intensities are in line with the other PAHs, the peak absorption intensity is therefore lower. The cause of the broadening does not appear to be insufficient cooling, since the width of nearby bands is limited by the laser bandwidth. Rather, we attribute this increased width to a blending of a large number of close-lying transitions. In contrast, theory only predicts one strong transition in this range. Another note must be made on the intensity of the CC modes intensity relative to the CH OOP bands: in general the CH OOP bands are attributed to neutral species whereas the CC modes in the 1200-1500 cm$^{-1}$ region are thought of as ionic in origin. Especially for pentacene, the strength of the CC modes is considerable (for neutrals in this case) compared to the CH OOP bands, which is not fully reflected in theory.

%                                     Two column figure (place early!)
%______________________________________________ Gamma_1 (lg rho, lg e)
   \begin{figure*}
   \centering

   \subfloat{\includegraphics[width=.4\textwidth]{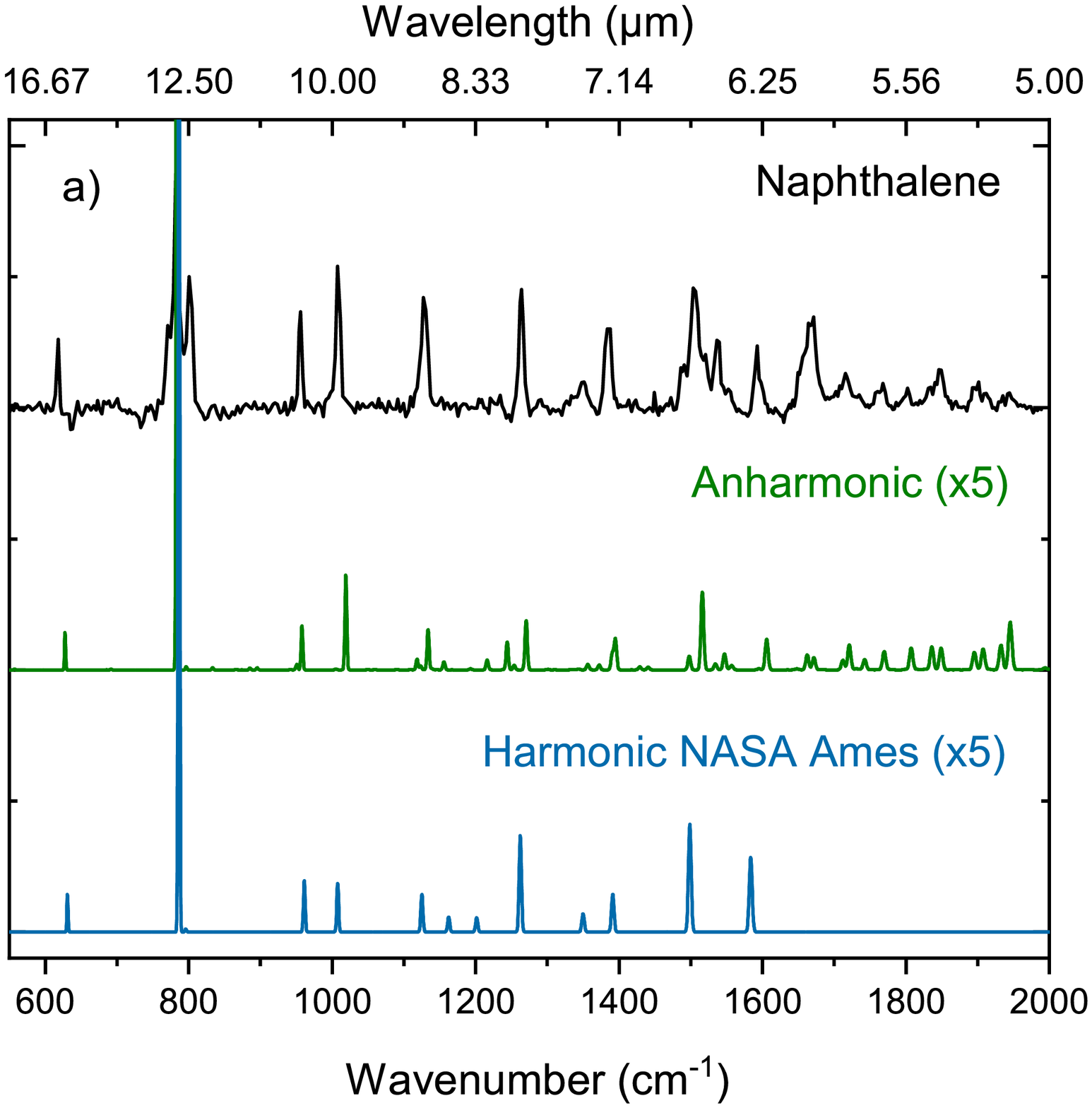}}
   \subfloat{\includegraphics[width=.4\textwidth]{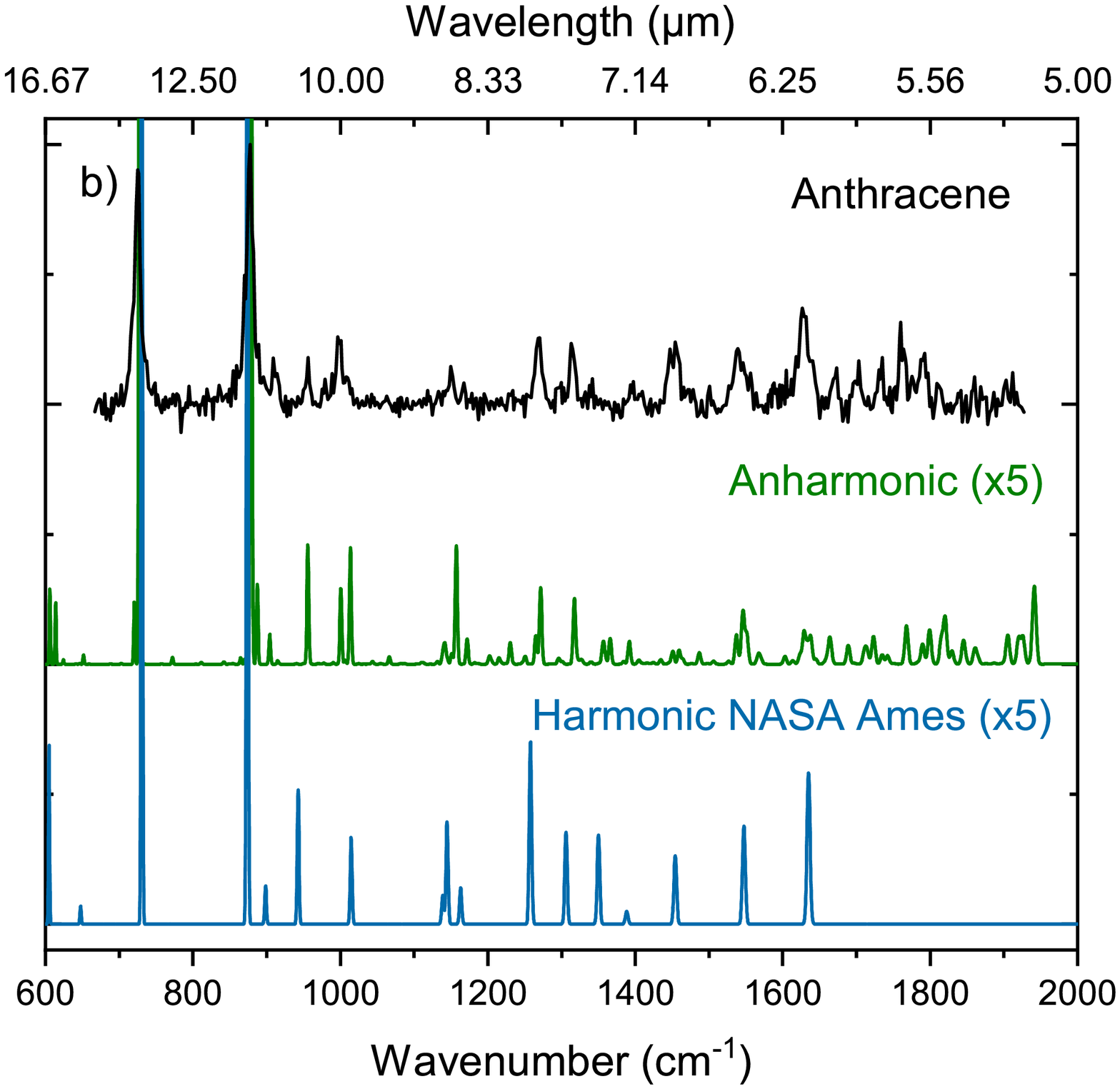}}\\
   \subfloat{\includegraphics[width=.4\textwidth]{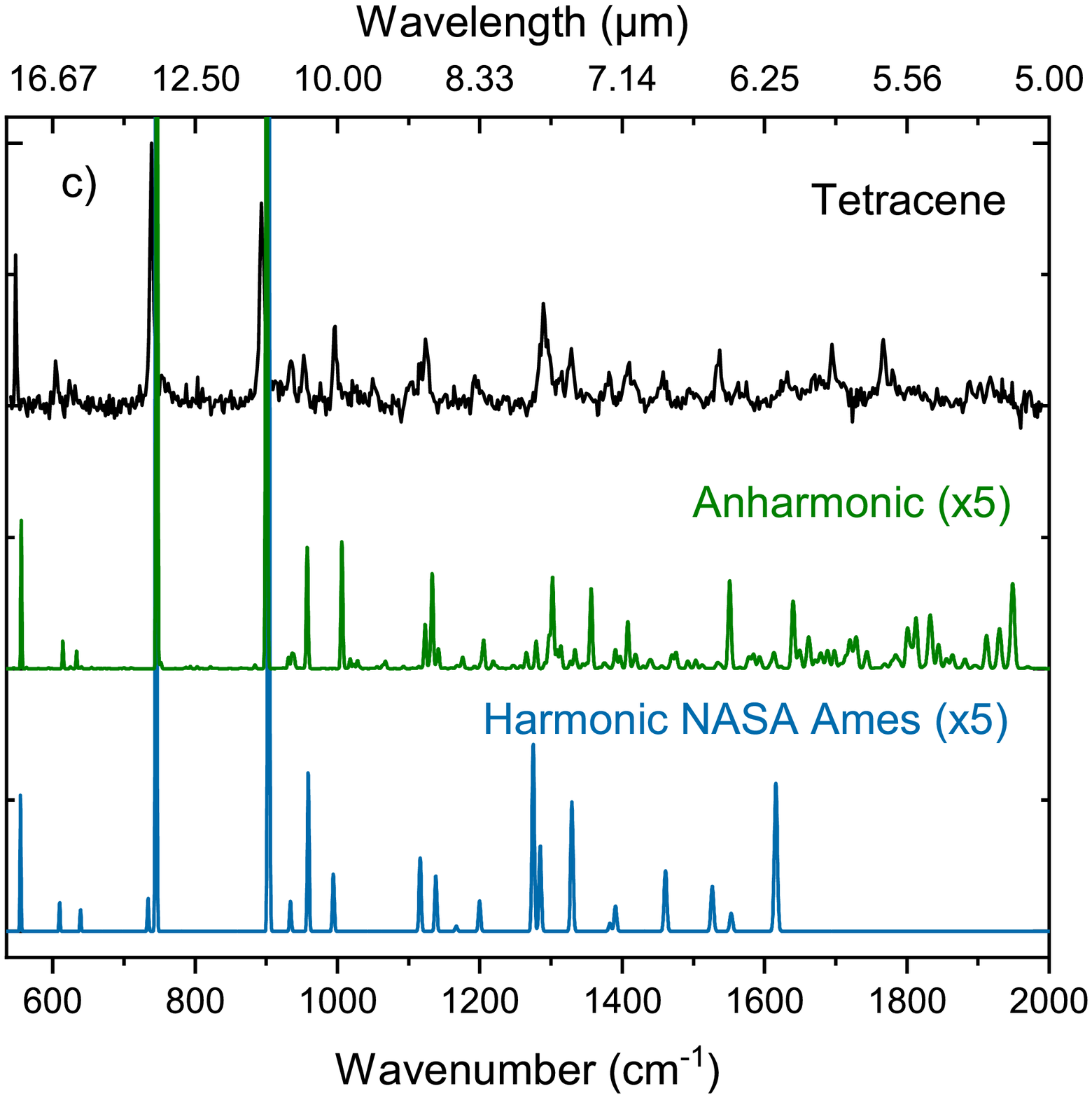}}\quad
   \subfloat{\includegraphics[width=.4\textwidth]{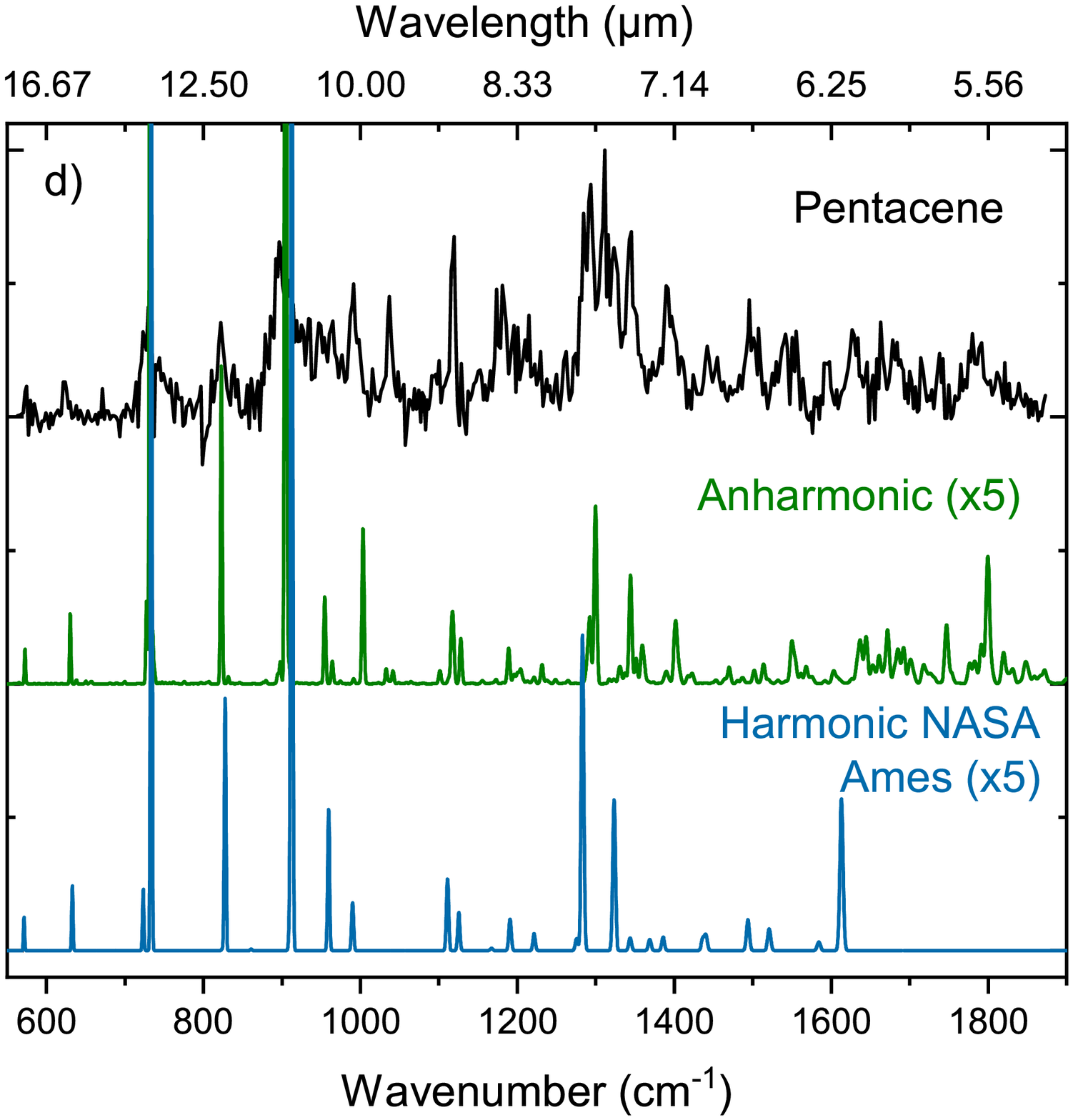}}

   \caption{Molecular beam gas-phase IR absorption spectra of (a) naphthalene, (b) anthracene, (c) tetracene and (d) pentacene (black). Predicted IR spectra using the harmonic and anharmonic approximation are given as blue and green traces, respectively. The  harmonic spectra have been obtained from the NASA Ames PAH spectral database \citep{Bauschlicher2018}. The predicted spectra are multiplied by a factor five, truncating the strongest bands to enable a better visualization of the low-intensity bands (normalized spectra can be found in the Appendix). }
              \label{FigGam1}%
    \end{figure*}
The most striking difference between the harmonic and anharmonic spectra is the activity in the 1600-2000 cm$^{-1}$ region. Below 1600 cm$^{-1}$, high-intensity bands are predicted reasonably well within the harmonic approximation (see Figure A.1), but at higher frequencies the activity observed in the experimental spectra can only be reproduced by calculations that account for anharmonicity (green traces). Such calculations show that these bands originate from combination bands built upon CH OOP bending modes. This assignment is in line with studies on substituted PAHs (see e.g., \citet{Bellamy1975} and references therein). In the case of pentacene, these bands contain a cumulative intensity of 69 km/mol, compared to a cumulative intensity of 350.5 km/mol of the region below 1600 cm$^{-1}$ according to the anharmonic calculations. Thus a significant amount of absorption (20\%) in the region between 1600 cm$^{-1}$ and 2000 cm$^{-1}$ is not predicted when using the harmonic approximation. 
The resolution of our experiments is limited by the laser bandwidth and thus does not allow us to fully resolve all combination bands in the 5-6 $\mu$m region. The larger polyacenes in particular show a high density of states in this region. However, based on the features that are resolved, standard deviations in frequencies of 0.4\%, 0.7\%, 0.5\%, and 0.4\% are found for naphthalene, anthracene, tetracene, and pentacene, respectively. A table containing the assigned transitions is provided in the Appendix (Table A.1-2). 

The agreement between theory and experiment is very good for naphthalene, both with respect to band position as well as shape. However, anthracene and tetracene show more differences between experiment and theory in the ‘combination band’ region. For anthracene, various distinct bands are measured between 1650 and 1800 cm$^{-1}$ whereas theory predicts an absorption feature that is more spread out. Another example is the rather strong band at 1767 cm$^{-1}$ in the experimental spectrum of tetracene whereas the calculated spectrum predicts a broader feature with less peak intensity. Apart from the broad absorption features of the CH OOP bands the spectrum of pentacene is overall in good agreement with theory. Particularly, the region between 1250 and 1400 cm$^{-1}$ changes significantly when anharmonicity is accounted for. Importantly, a comparison with the matrix isolation spectrum reported by \citet{Hudgins1998}, which has been used so far in astronomical models, shows several differences. One of the main differences concerns the intensities of the bands at 1288 and 1349 cm$^{-1}$ , which have about equal intensity in our experiment, whereas in the MIS spectrum they differ by about a factor of 20. A comparison of the matrix isolation spectra to our results is shown in the Appendix (Figure A.2). 

The excellent agreement between experiment and theory allows us to use the theoretical calculations with confidence to investigate how the size of the polyacene correlates with the ratio of the integrated intensity in the 1600-2000 cm$^{-1}$ combination band region and the integrated intensity of fundamental bands (see Figure 2). Figure  2 shows that upon increasing the size of the PAH, combination bands acquire a larger absolute and fractional part of the total IR absorption intensity, which is in line with the higher density of vibrational states in  larger PAHs. The fractional increase is observed both relative to the CH OOP bands as well as the integrated intensity of the bands below 1600 cm$^{-1}$. These ratios could thus provide a sensitive means to determine the size of (linear) PAHs in interstellar clouds. Remarkable in this respect is that the relation found by \citet{Boersma2009} for a larger ensemble of PAHs is opposite to the trend found in this study for polyacenes: for a larger number of carbon atoms, the combination bands/CH OOP ratio decreases. This may simply be explained by the fact that in this study a particular subset of PAHs is taken into account rather than a diverse ensemble. Another interesting observation that can be made concerns the general appearance of the 5-6 $\mu$m region. Only two distinct bands belonging to the AIBs appear in this region, at 5.25 and 5.7 $\mu$m \citep{Boersma2009}. The present study shows that this region is largely built upon combination bands involving the CH OOP modes which, as discussed in the beginning of this section, are very sensitive to the periphery of the molecule in question. The presence of only two features in the AIB bands therefore strongly suggests a large similarity in the periphery of the PAHs present in these regions since a variety of peripheries would lead to a more uniform absorption in the combination band region. 
%                                     Two column figure (place early!)
%______________________________________________ Gamma_1 (lg rho, lg e)
   \begin{figure*}
   \centering
  \includegraphics[width=.4\textwidth]{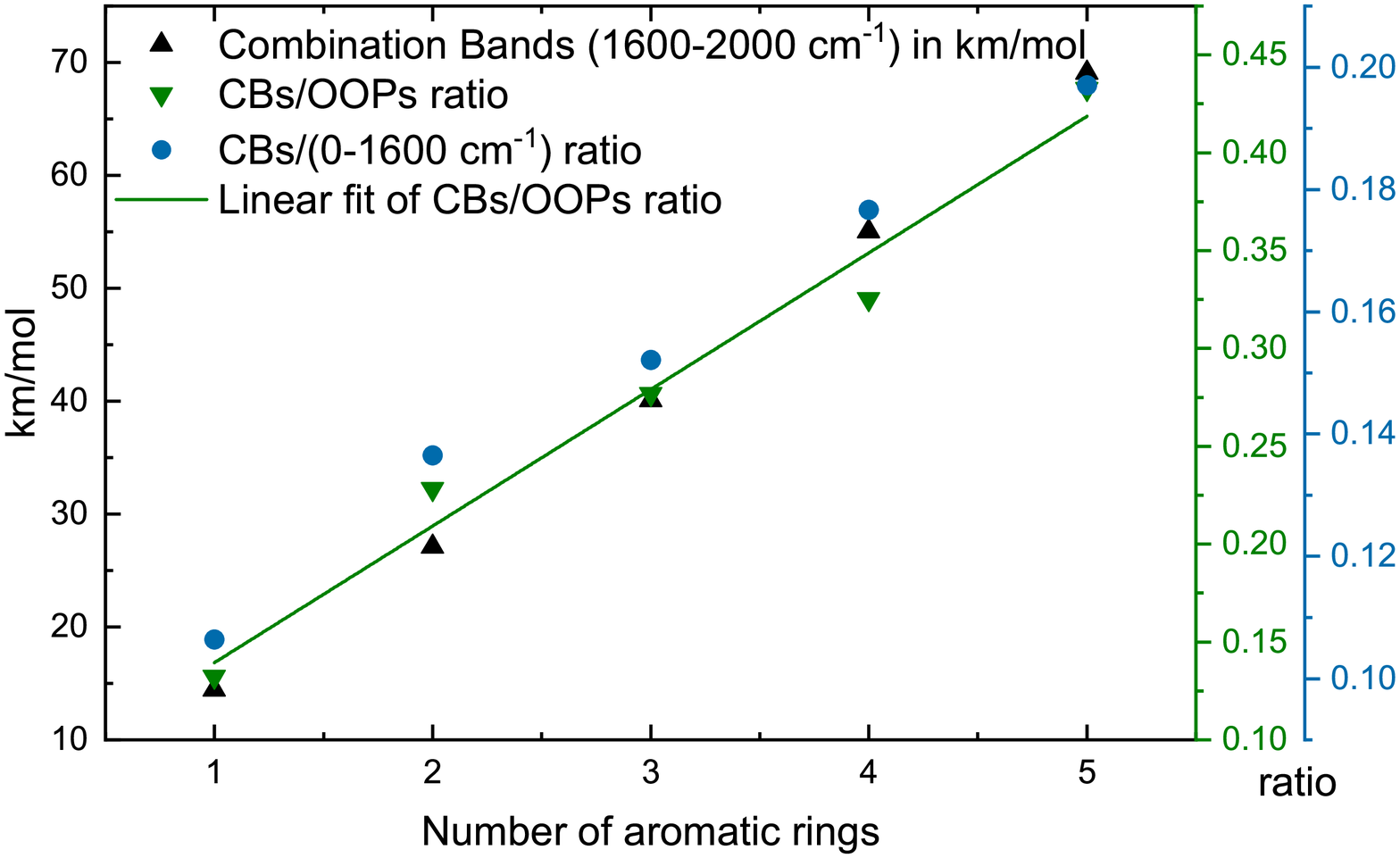}
   \caption{Cumulative intensity of the combination band (CB) region (1600-2000 cm$^{-1}$, black) and the ratio to the 0-1600 cm$^{-1}$ region (blue) as well as to the CH out-of-plane (OOPs) bands (green) as a function of the number of aromatic rings in polyacenes (and benzene for the single ring species). The green trace corresponds to a linear fit to the CBs/OOPs ratio data with a slope of 0.0698 $\pm$ 0.0062 and an intercept of 0.070 $\pm$ 0.021. }
              \label{FigGam2}%
    \end{figure*}
    
Another region that is key to finding signatures directly related to the size of PAHs is the far-IR region \citep{Boersma2010,Ricca2010,Zhang1996}. This region is prone to be susceptible to the effects of anharmonicity, but has so far proven to be difficult to study, certainly under conditions relevant for astronomical observations. In the far-IR region, large-amplitude vibrations such as the anharmonic ring puckering motion of PAHs are active. Particularly for such large-amplitude modes, the anharmonic VPT2 calculations may lead to inaccurate results since the truncation of the Taylor expansion of the potential energy curve in VPT2 is an oversimplification and higher-order terms need to be taken into account \citep{Csaszar2012}. Moreover, extreme caution must be taken when applying VPT2 to modes where the harmonic term does not dominate the shape of the potential energy curve \citep{Barone2014,Mahe2015}. For example, in the case of the hetero-PAH aminophenol it was shown previously that ring puckering is poorly described by the VPT2 treatment \citep{Laane1967,Yatsyna2016}. The anharmonic bands in the mid-IR provide such a way to assess the low-energy part of the potential energy surface as they involve combination bands built upon one or more fundamentals from the far-IR. For example, in the case of naphthalene, a broader band is present at 1351 cm$^{-1}$ in both experiment and anharmonic calculations. This band has contributions from an IR-active fundamental band at 1372 cm$^{-1}$ and a combination band of the IR-inactive 387 and 970 cm$^{-1}$ modes. The good agreement between experiment and theory thus indicates that the calculated mode at 387 cm$^{-1}$ is well described. Similarly, we find for tetracene that several bands between 996 and 1123 cm$^{-1}$ involving far-IR fundamentals are well predicted in the anharmonic calculations. On the other hand, in naphthalene a 479 and 772 cm$^{-1}$ combination band is predicted at 1244 cm$^{-1}$ but is not observed in the experimental spectrum. Also for anthracene, some predicted bands are not present in the experiment: the predicted band at 614 cm$^{-1}$ is missing in the anthracene MIS spectrum \citep{Hudgins1998a} and several combination bands are not present in our gas phase spectrum around 1700 cm$^{-1}$. Although the present study provides some insight into the validity of calculations of the far-IR region, it is clear that further far-IR studies -- and in particular experimental studies in the region itself -- are needed for a more detailed characterization. Such studies will indeed shortly be performed in our laboratories.

Our studies have shown for the larger part excellent agreement between experimental and anharmonic theoretical spectra. However, they also show that higher-order perturbations in the Taylor expansion of the PES or dipole moment surface, which are as yet not incorporated into the theory, do play a role. This is most evident from the intense absorption feature at 783 cm$^{-1}$ in the spectrum of naphthalene. Previously, this feature was observed as a doublet in both matrix isolation \citep{Hudgins1998a} and in cationic form in supersonically cooled gas phase \citep{Piest1999}. With the present resolution it becomes clear that this feature actually consists of three well-resolved bands at 770.8, 782.7, and 800.5 cm$^{-1}$ (Figure 3). In a study at higher temperatures \citep{Pirali2009}, the triplet band structure is not observed, but clear rotational structure is shown. The side bands in our observation cannot be rotational contours, since the rotational temperature in our experiments is much lower than would be required for the given splitting. Our present anharmonic calculations that only take double combination bands into account do not reproduce this splitting. One could therefore hypothesize that triple combination bands might be the origin of the splitting. This appears to be contradicted by the studies reported by \citet{Bloino2015} where transitions with three quanta in the final state are calculated. These calculations do not show intense transitions at 770.8 and 800.5 cm$^{-1}$ but rather in the near-IR region. Inspection of the vibrational levels at this energy shows that the state density is high enough to support the presence of three bands. At the same time we have to conclude that the calculation of their intensities is apparently still far from what it should be. A contributing factor might be that for the involved motions the dipole moment surface is not well described by the calculations. Studies focusing on this dipole moment surface are therefore of interest and are being pursued.  
%                                     Two column figure (place early!)
%______________________________________________ Gamma_1 (lg rho, lg e)
   \begin{figure*}
   \centering
  \includegraphics[width=.4\textwidth]{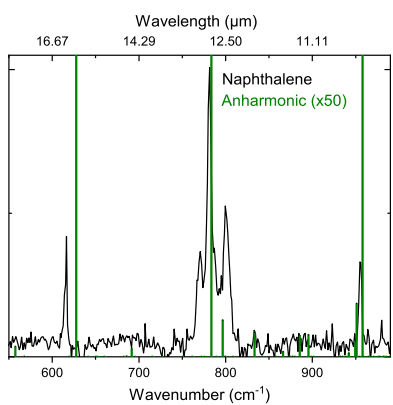}
   \caption{Zoom-in of the gas-phase IR absorption spectrum of jet-cooled naphthalene (black). The predicted absorption spectrum using the anharmonic approximation is shown in green. The simulation is multiplied by a factor 50 for better visualization of low-intensity bands.}
              \label{FigGam3}%
    \end{figure*}

\section{Astrophysical implications}

The present study confirms the far-reaching consequences of anharmonicity in the mid-IR spectra of linear PAHs. firstly, it demonstrates how anharmonic calculations are a sine qua non condition to simulate the 5-6 $\mu$m region. The combination bands that underlie the observed activity in this region contain a significant amount of IR absorption intensity, and this fraction increases with the size of the polyacene. In fact, the PAHs studied here suggest a linear relationship between the length of the polyacene and the intensity ratio of combination bands to fundamentals. A surprising aspect of the measured spectra is that, while there are only one or two strong CH OOP modes, the 1600-2000 cm$^{-1}$  region is rich in spectral detail. The strong CH OOP modes couple to more than one IR inactive mode, leading to this rich spectrum. Interestingly, the interstellar spectra are dominated by only two bands. From this it is clear that further study of the 
relative intensities of larger PAHs  is important, since deviating intensities in calculations can lead to misinterpretation of astronomical data. Secondly, in addition to the combination bands, the length of the polyacene also has a profound influence on the overall appearance of the IR spectrum: the OOP CH bending modes are a sensitive probe for the periphery of the molecule and since the combination bands are mostly comprised of at least one CH OOP, this also holds for the combination bands. Thirdly, anharmonic effects well below 1000 cm$^{-1}$ have been observed -- and naphthalene provides in this respect a most prominent example -- that cannot be neglected since bands with a very strong absorption are involved. The present theoretical approach is not capable of providing a consistent explanation for the observed activity, and we thus conclude that higher-order perturbations associated with both in the potential energy surface and the dipole moment surface are more important than previously assumed. The resulting intensity discrepancies can have important consequences in the context of the analysis of astronomical observations and lead to an incorrect interpretation in terms of relative abundances of charged or neutral species. Finally, the observations of combination bands built upon far-IR modes give a preliminary indication of the accuracy of the anharmonic calculations in predicting far-IR modes which are important as they have the potential to be sensitive probes of PAH size \citep{Boersma2010}. Our results suggest that the far-IR frequencies are accurately predicted within a few wavenumbers, but their coupling and intensities still need further improvement.

\section{Conclusions}

High-resolution mid-IR absorption spectra of the polyacenes naphthalene, anthracene, tetracene, and pentacene were obtained using IR-UV double resonance laser spectroscopy under astronomically relevant conditions. Comparison of these spectra with theoretical spectra predicted by harmonic and anharmonic calculations unambiguously shows the necessity to incorporate anharmonicity into astronomical models. Our anharmonically predicted spectra have a frequency deviation of 0.5\% without the use of a scaling factor. We show that the 5-6 $\mu$m region contains a significant fraction of intensity associated with combination bands that should not be ignored in the analysis of astronomical spectra. For the polyacenes series studied here, the fraction of combination band intensity scales linearly with the length of the polyacene. This underlines the importance of incorporating anharmonicity in spectral analyses since PAHs in the interstellar medium are generally thought to be much larger than pentacene. The assignment of several combination bands has also provided a preliminary indication of the accuracy of current computational methodologies for calculating frequencies of far-IR bands. These far-IR bands are of interest since they could serve as a ‘label’ for  the substructure of PAHs, and are currently being pursued in our laboratories. 

Overall, the present anharmonic calculations lead to very good agreement with experimental results. However, detailed analyses of, for example, the out-of-plane CH bending region of naphthalene show that, in particular regions, considerable differences 
still occur. Since these regions are often leading in astronomical analyses, further studies are required to determine the cause of these differences. One of the aspects which in that respect needs further attention is the dipole moment surface, which so far has been assumed to be constant near the equilibrium geometry. 

In the present study, pentacene needed to be studied using laser desorption instead of heating which up until now could be used to seed molecules into the molecular beam. Our IR absorption studies show that spectra obtained with laser desorption compare favorably with spectra obtained under heating conditions. This opens up a route to investigate the spectral signatures of large PAHs at the same level of detail as has been shown possible for smaller PAHs. Preliminary experiments performed in our laboratories on much larger PAHs indeed support this conclusion.

\begin{acknowledgements}
This work was supported by The Netherlands Organization for Scientific Research (NWO) and is part of the Dutch Astrochemistry Network (DAN) II. We gratefully acknowledge the NWO for the support of the FELIX Laboratory. AC is supported by an NWO VENI grant (No. 639.041.543). The theoretical work was performed on the Dutch national e-infrastructure with the support of SURFsara. Furthermore, we would like to thank the FELIX laboratory team for their experimental assistance.

\end{acknowledgements}

\bibliographystyle{aa}
\bibliography{bieb.bib}

\onecolumn
\pagebreak
\begin{appendix}

\section{ }

   \begin{figure*}[hbtp]
   \centering

   \subfloat{\includegraphics[width=.4\textwidth]{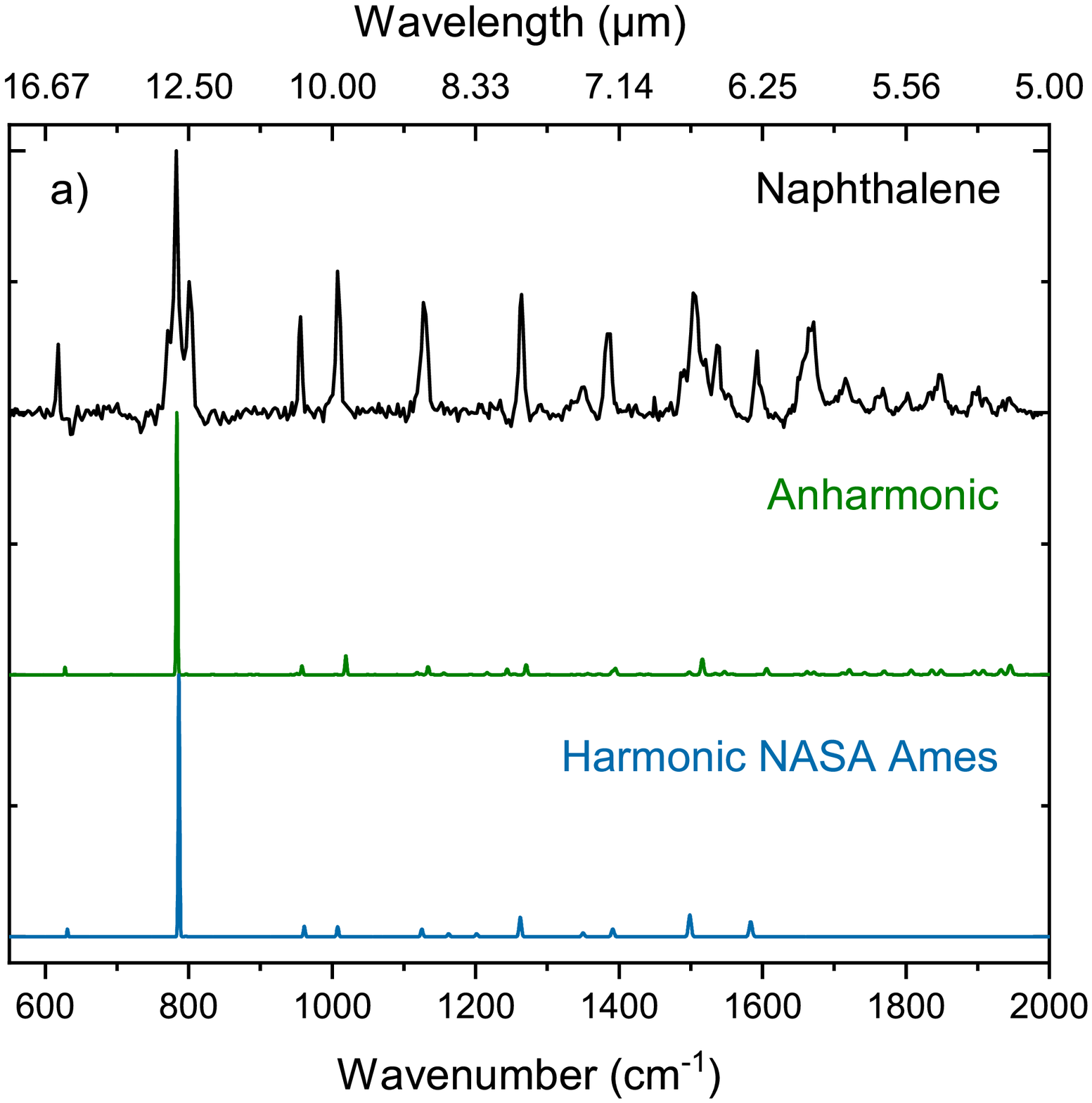}}\quad
   \subfloat{\includegraphics[width=.4\textwidth]{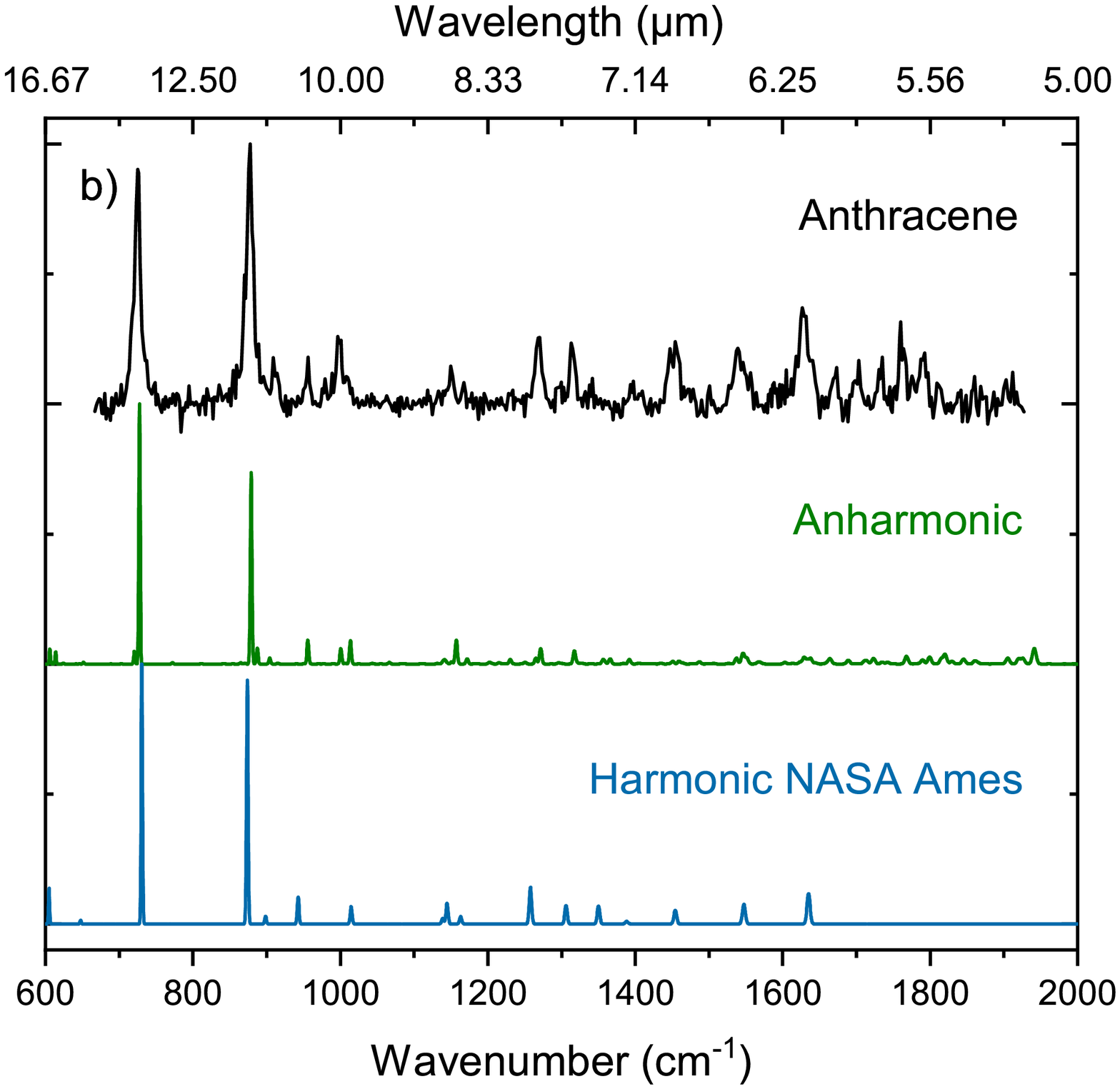}}\\
   \subfloat{\includegraphics[width=.4\textwidth]{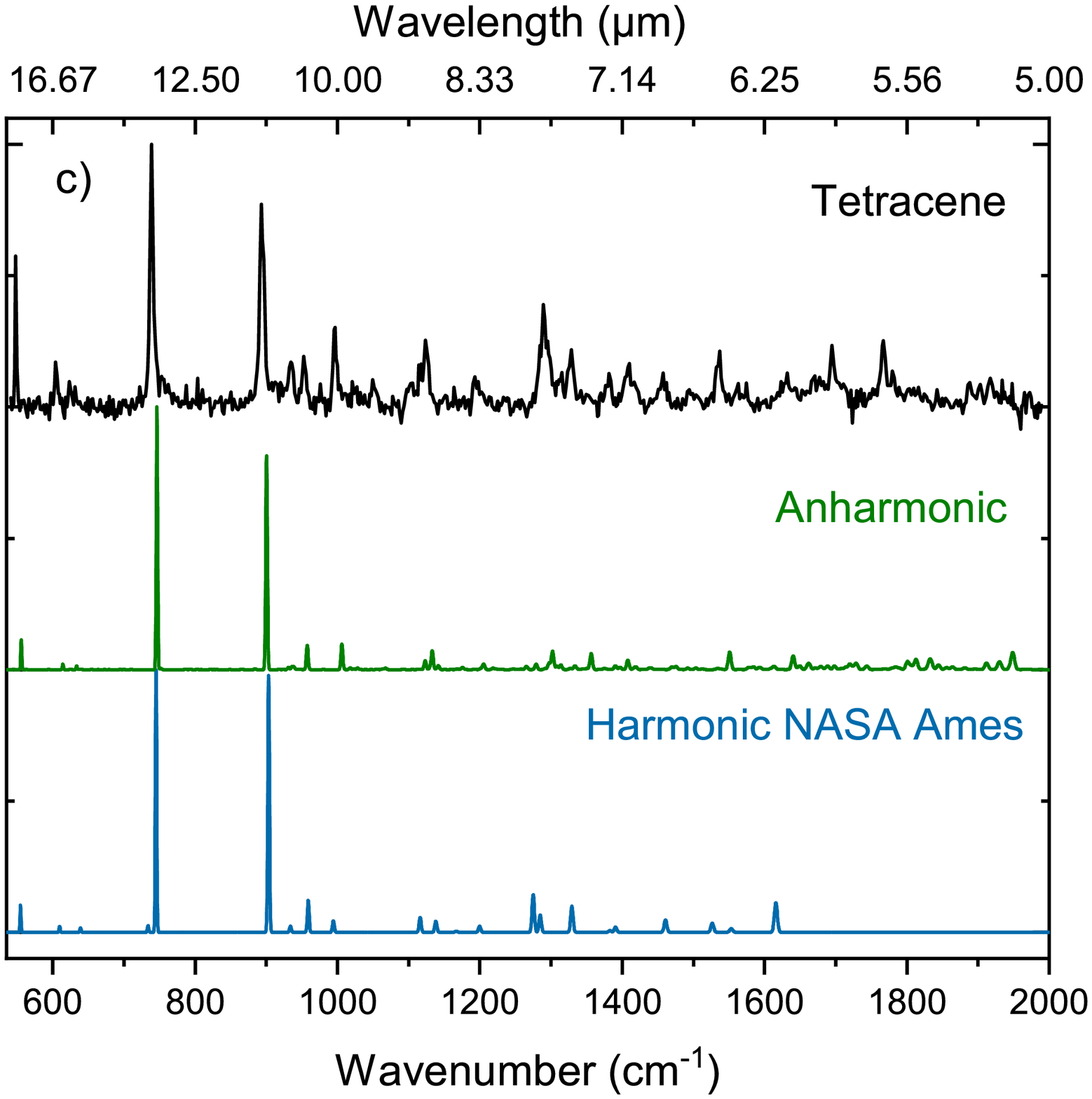}}\quad
   \subfloat{\includegraphics[width=.4\textwidth]{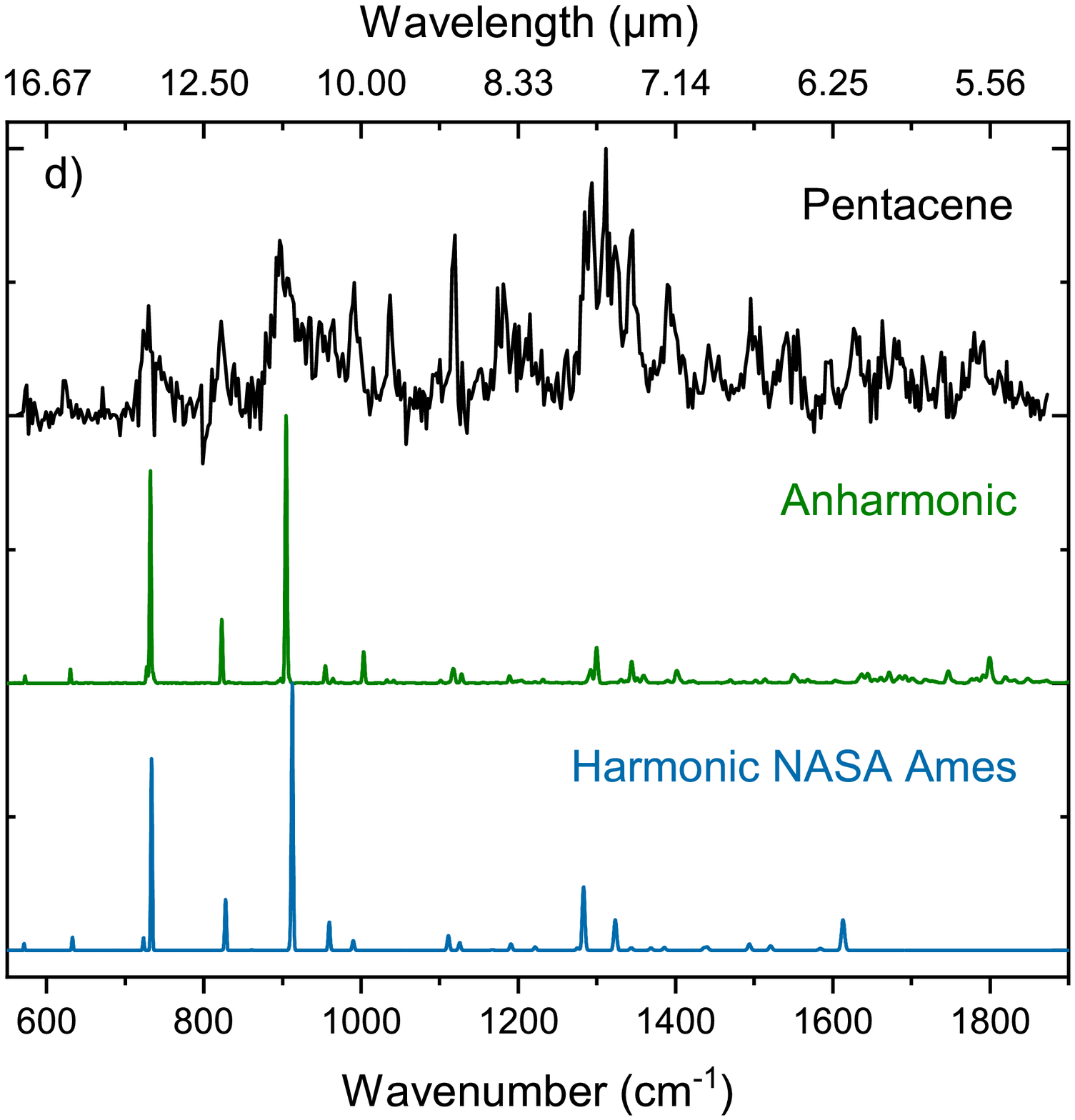}}

   \caption{Molecular beam gas-phase IR absorption spectra of (a) naphthalene, (b) anthracene, (c) tetracene, and (d) pentacene in black. Normalized predicted IR spectra using the harmonic and anharmonic approximation are in blue and green, respectively. The  harmonic spectra are obtained from the NASA Ames PAH spectral database\citep{Bauschlicher2018}. }
              \label{FigGam4}%
    \end{figure*}

   \begin{figure*}
   \centering

   \subfloat{\includegraphics[width=.4\textwidth]{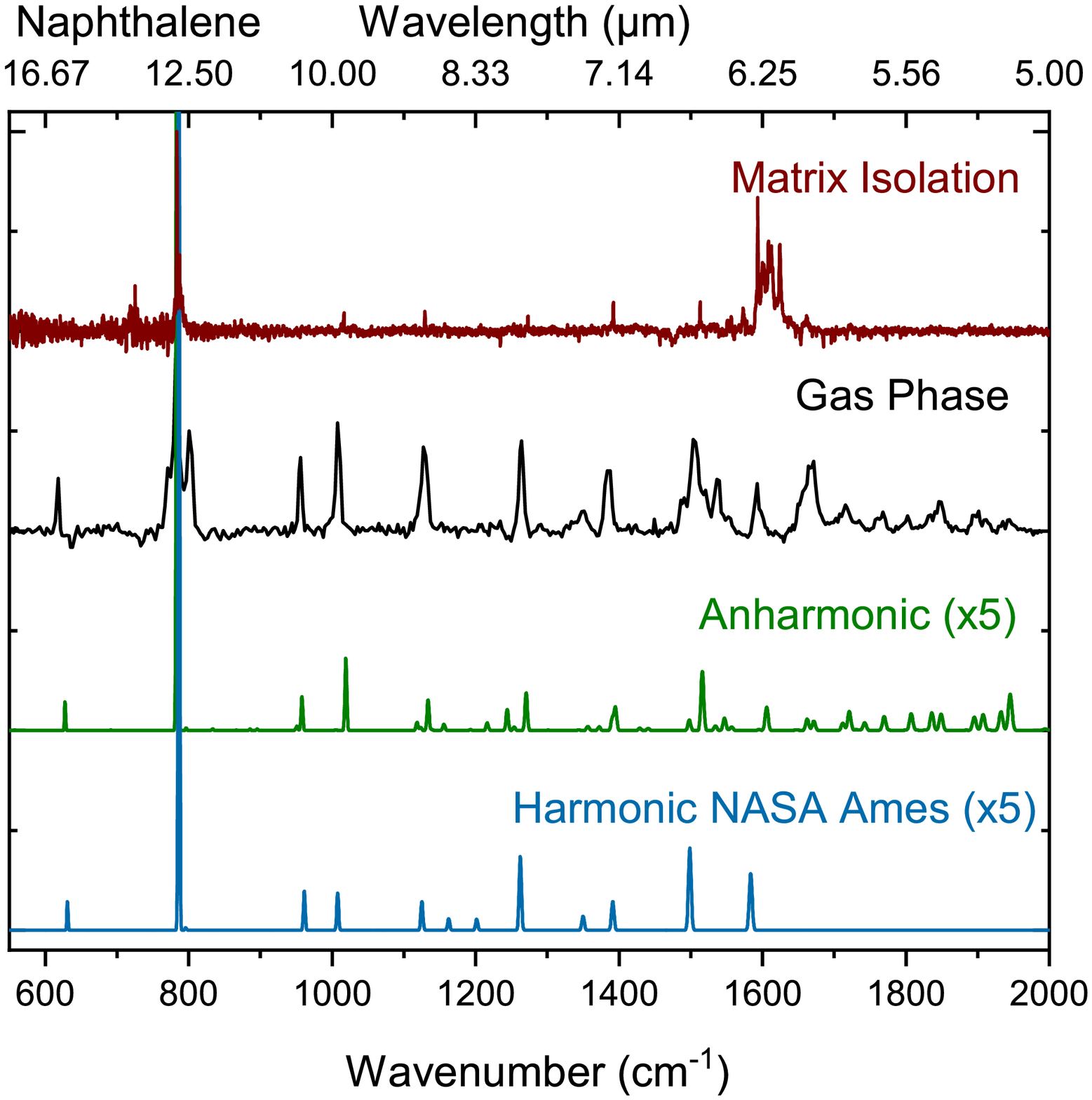}}\quad
   \subfloat{\includegraphics[width=.4\textwidth]{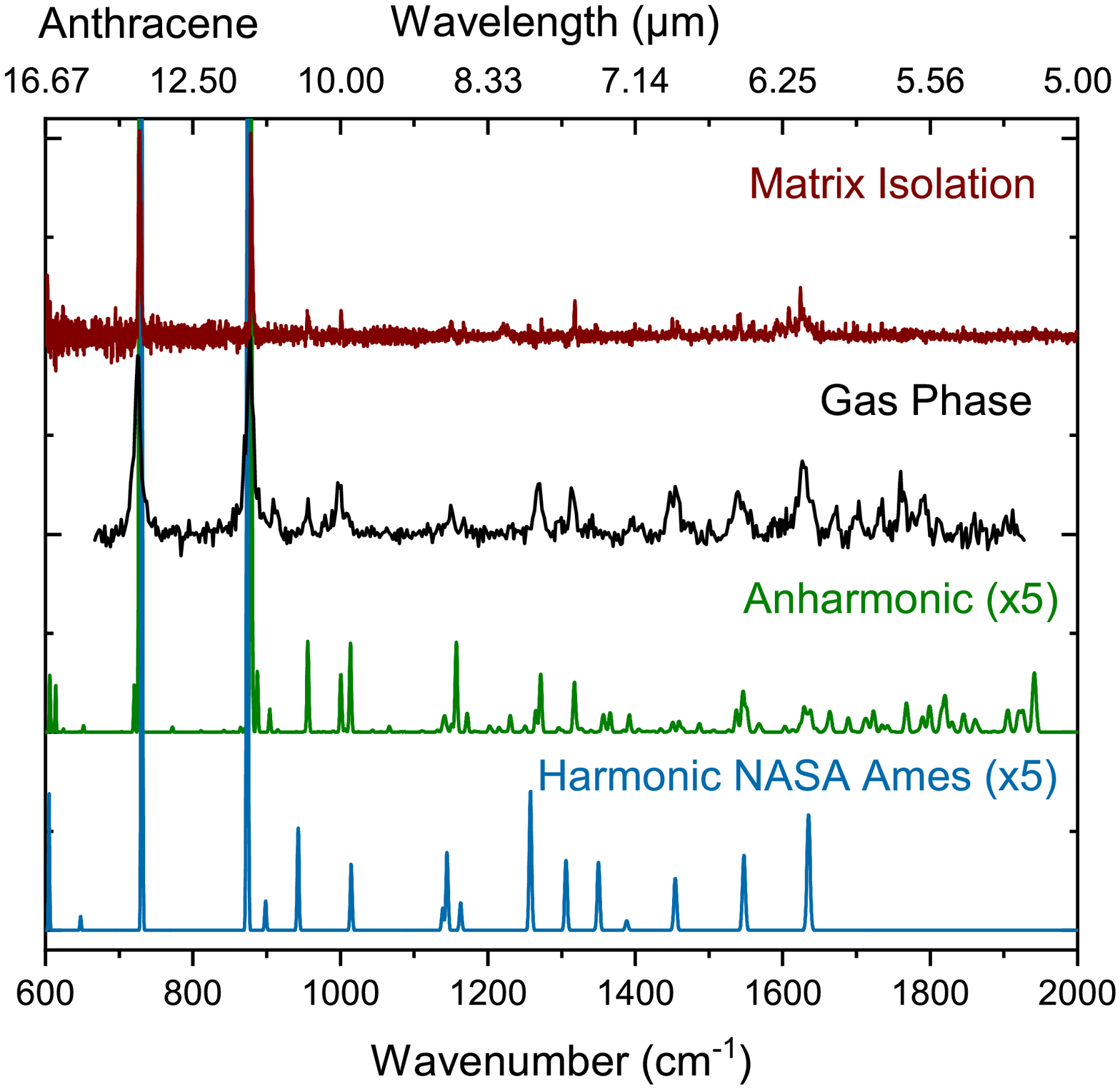}}\\
   \subfloat{\includegraphics[width=.4\textwidth]{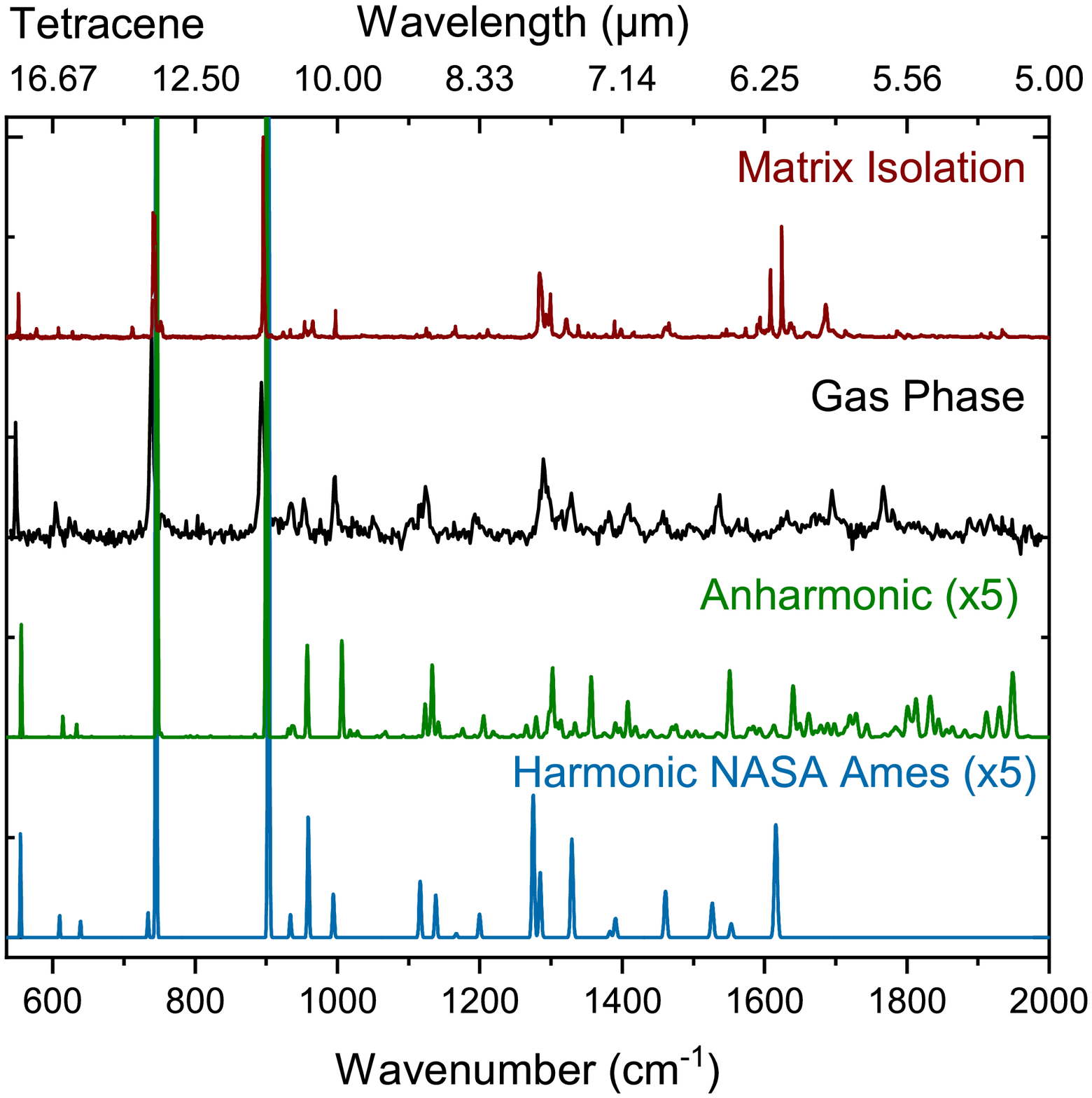}}\quad
   \subfloat{\includegraphics[width=.4\textwidth]{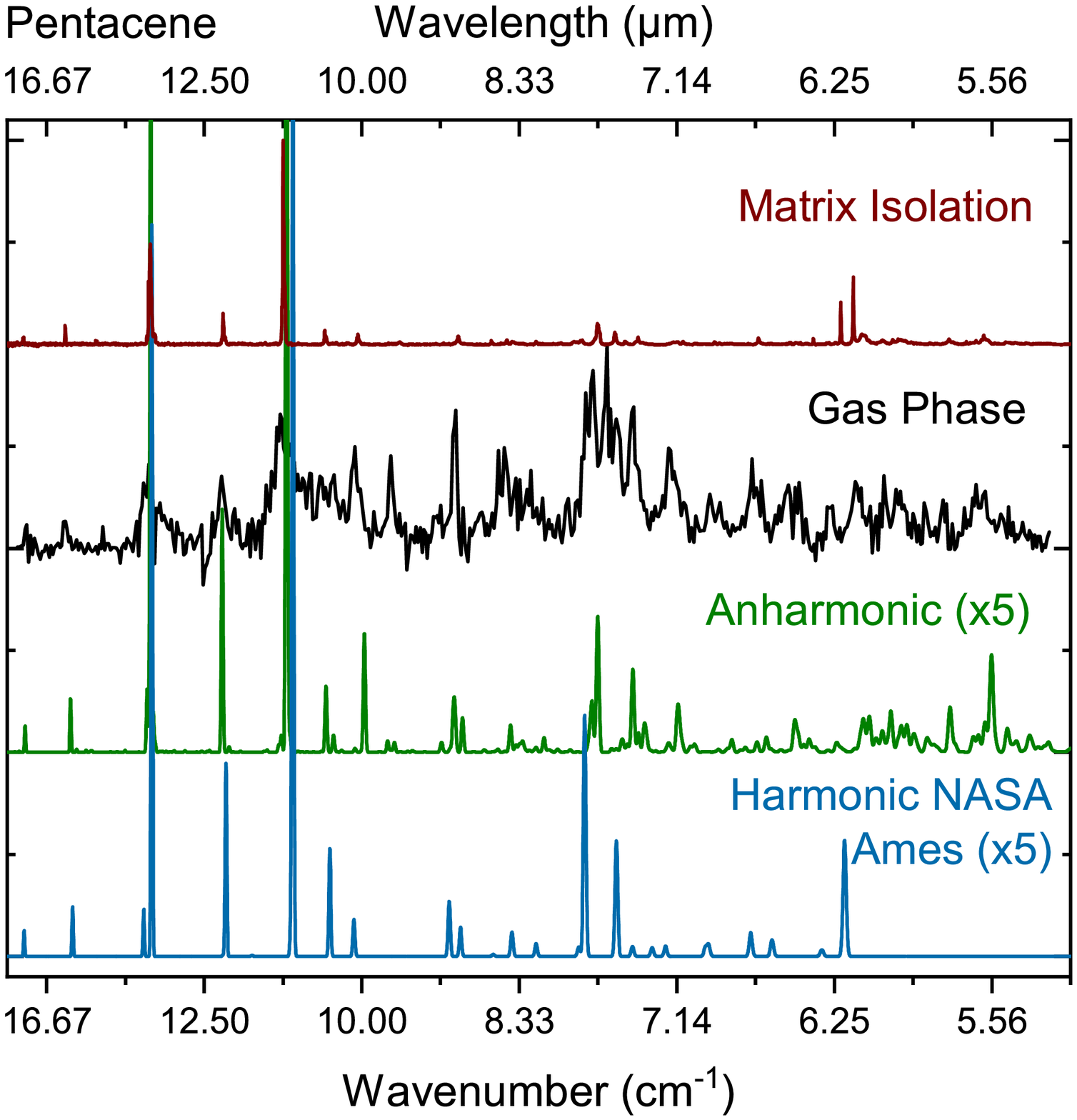}}

   \caption{Matrix Isolation and molecular beam gas-phase IR absorption spectra of naphthalene, anthracene, tetracene, and pentacene in red and black, respectively. The MIS spectra of naphthalene and anthracene were obtained from \citet{Roser2010,Roser2014}, while the MIS spectra of tetracene and pentacene were obtained from the NASA Ames PAH spectral database \citet{Bauschlicher2018}. Predicted IR spectra using the harmonic and anharmonic approximation are in blue and green, respectively. The  harmonic spectra were obtained from the NASA Ames PAH spectral database \citep{Bauschlicher2018}.}
              \label{FigGam5}%
    \end{figure*}
   \begin{figure*}
   \centering
  \includegraphics[width=.6\textwidth]{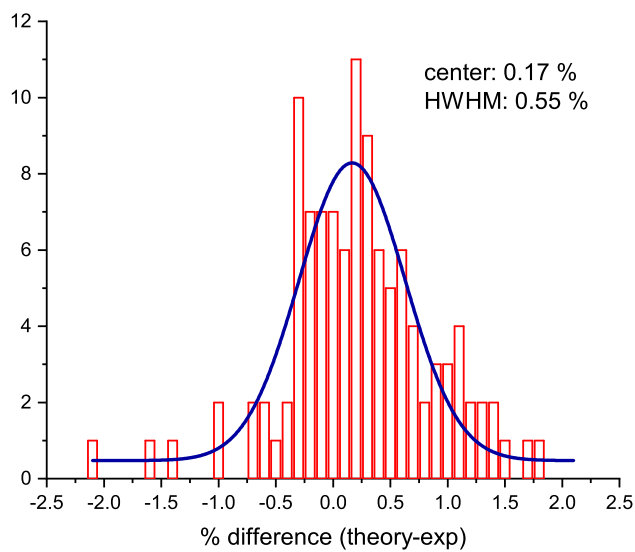}
   \caption{Histogram showing the percent differences in line position between anharmonic theory and experiment for naphthalene, anthracene, tetracene, and pentacene in the region measured in this study ($\pm$ 2000-550 cm$^{-1}$)}
              \label{FigGam6}%
    \end{figure*}

\begin{table}
\caption{Measured bands of naphthalene and anthracene  with frequencies in (cm$^{-1}$) and normalized intensities.}\label{table1}
\centering
\begin{tabular}{lrlrlrlr}
\hline\hline

\multicolumn{4}{c}{Naphthalene (C$_{10}$H$_{12}$)}&\multicolumn{4}{c}{Anthracene (C$_{14}$H$_{12}$)}\\

exp freq.& rel. int.& calc. freq.& rel. int.&exp freq.& rel. int.& calc. freq.& rel. int. \\
\hline
618.24 & 0.11 & 627.4  & 0.05 & 724.8  & 0.84 & 727.6  & 1.00  \\
770.8  & 0.30 & …      & …    & 877.6  & 1.00 & 879.0  & 0.60  \\
782.7  & 1.00 & 784.6  & 1.00 & 908.8  & 0.25 & 904.2  & 0.03  \\
800.5  & 0.54 & 795.9  & 0.00 & 956.3  & 0.03 & 955.8  & 0.09  \\
956.0  & 0.24 & 958.5  & 0.02 & 979.0  & 0.15 & 1000.0 & 0.04  \\
1007.7 & 0.53 & 1018.0 & 0.04 & 996.2  & 0.49 & 1013.0 & 0.02  \\
1127.4 & 0.56 & 1133.2 & 0.03 & 1149.0 & 0.20 & 1157.2 & 0.02  \\
1264.0 & 0.41 & 1269.0 & 0.02 & 1167.4 & 0.09 & 1171.9 & 0.02  \\
1351.2 & 0.20 & 1350.0 & 0.01 & 1233.0 & 0.05 & 1230.0 & 0.01  \\
1387.9 & 0.41 & 1388.0 & 0.02 & 1270.0 & 0.04 & 1271.0 & 0.02  \\
1491.1 & 0.17 & 1496.0 & 0.00 & 1299.6 & 0.10 & 1300.9 & 0.00  \\
1503.6 & 0.91 & 1503.0 & 0.00 & 1313.1 & 0.33 & 1317.4 & 0.07  \\
1536.5 & 0.40 & 1537.0 & 0.03 & 1328.0 &      & 1327.0 &       \\
1595.0 & 0.22 & 1606.0 & 0.01 & 1342.1 & 0.09 & 1356.5 & 0.02  \\
1671.1 & 0.87 & 1673.0 & 0.01 & …      & …    & 1365.9 & 0.01  \\
1717.0 & 0.34 & 1722.0 & 0.01 & 1397.4 & 0.08 & 1391.9 & 0.00  \\
1768.1 & 0.14 & 1766.0 & 0.00 & 1409.6 & 0.04 & 1405.0 & 0.00  \\
1802.1 & 0.10 & 1808.0 & 0.01 & 1447.1 & 0.24 & 1451.1 & 0.00  \\
1849.3 & 0.33 & 1850.0 & 0.01 & 1454.5 & 0.30 & 1459.3 & 0.01  \\
1898.0 & 0.18 & 1896.0 & 0.01 & 1538.9 & 0.60 & 1543.0 & 0.01  \\
1944.7 & 0.11 & 1947.0 & 0.03 & 1626.0 & 0.98 & 1633.0 & 0.02  \\
       &      &        &      & 1673.0 & 0.14 & 1664.0 & 0.03  \\
       &      &        &      & 1703.0 & 0.19 & 1689.0 & 0.01  \\
       &      &        &      & 1735.0 & 0.18 & 1723.0 & 0.00  \\
       &      &        &      & 1760.0 & 0.41 & 1768.0 & 0.00  \\
       &      &        &      & 1775.0 & 0.14 & 1789.0 & 0.00  \\
       &      &        &      & 1792.0 & 0.34 & 1799.0 & 0.00  \\
       &      &        &      & 1812.0 & 0.06 & 1815.0 & 0.02 \\
       
       \hline
\end{tabular}
\end{table}

\begin{table}
\caption{Measured bands of tetracene and pentacene  with frequencies in (cm$^{-1}$) and normalized intensities.}\label{table2}

\centering
       \begin{tabular}{lrlrlrlr}
\hline\hline

\multicolumn{4}{c}{Tetracene (C$_{10}$H$_{12}$)}&\multicolumn{4}{c}{Pentacene (C$_{14}$H$_{12}$)}\\

exp freq.& rel. int.& calc. freq.& rel. int.&exp freq.& rel. int.& calc. freq.& rel. int. \\
\hline
       
548.0  & 0.19 & 555.6  & 0.11 & 574.5  & 0.02 & 573.0  & 0.02  \\
604.0  & 0.11 & 614.1  & 0.02 & 621.2  & 0.04 & 630.0  & 0.04  \\
623.0  & 0.05 & 624.6  & 0.00 & 730.2  & 0.36 & 732.0  & 0.71  \\
631.3  & 0.04 & 633.4  & 0.01 & 823.3  & 0.22 & 823.0  & 0.22  \\
738.7  & 1.00 & 746.3  & 1.00 & 897.9  & 1.00 & 905.0  & 1.00  \\
892.9  & 0.96 & 900.0  & 0.81 & 948.9  & …    & 955.0  & …     \\
934.4  & 0.21 & 938.7  & 0.01 & 967.0  & 0.41 & 964.0  & 0.09  \\
952.4  & 0.18 & 957.4  & 0.09 & 993.5  & 0.32 & 1003.0 & 0.13  \\
975.8  & 0.06 & …      & …    & 1017.4 & …    & …      & …     \\
996.6  & 0.29 & 1006.0 & 0.09 & 1039.4 & 0.22 & 1033.0 & 0.03  \\
1049.4 & 0.10 & 1067.3 & 0.01 & 1103.0 & 0.12 & 1116.0 & 0.02  \\
1123.8 & 0.35 & 1133.0 & 0.07 & 1122.3 & 0.26 & 1128.0 & 0.11  \\
1081.3 & 0.11 & 1092.6 & 0.00 & 1155.6 & 0.11 & …      & …     \\
1195.4 & 0.16 & 1205.5 & 0.02 & 1176.8 & 0.30 & 1189.0 & 0.05  \\
1289.1 & 0.62 & 1302.3 & 0.07 & 1183.9 & 0.18 & …      & …     \\
1328.6 & 0.42 & 1333.4 & 0.01 & 1204.0 & 0.16 & 1204.0 & 0.03  \\
1381.7 & 0.17 & 1390.5 & 0.01 & 1218.0 & 0.10 & …      & …     \\
1410.0 & 0.32 & 1407.9 & 0.04 & 1232.8 & …    & …      & …     \\
1457.7 & 0.18 & 1475.6 & 0.01 & 1265.6 & …    & …      & …     \\
1498.6 & 0.13 & 1503.4 & 0.01 & 1288.0 & 0.67 & 1292.0 & 0.28  \\
1536.9 & 0.29 & 1551.1 & 0.05 & 1297.3 & …    & 1299.0 & …     \\
1563.4 & 0.08 & 1584.4 & 0.01 & 1315.0 & …    & 1317.0 & …     \\
1574.4 & 0.08 & 1592.2 & 0.01 & 1319.6 & 0.74 & 1322.0 & …     \\
1632.1 & 0.28 & 1640.3 & 0.05 & 1326.7 & …    & 1324.0 & …     \\
1670.6 & 0.33 & …      & …    & 1349.1 & 0.47 & 1344.0 & 0.24  \\
1694.5 & 0.48 & 1698.2 & 0.01 & 1393.9 & 0.48 & 1402.0 & 0.14  \\
1767.0 & 0.47 & …      & …    & 1446.2 & …    & 1453.0 & …     \\
1779.6 & …    & 1784.0 & 0.01 & 1459.1 & 0.27 & 1465.0 & 0.04  \\
1809.5 & 0.21 & 1831.6 & 0.03 & 1500.0 & 0.41 & 1502.0 & 0.05  \\
1885.7 & …    & …      & ...  & 1512.0 & …    & 1514.0 & …     \\
1902.9 & 0.20 & 1911.9 & 0.02 & 1546.9 & …    & 1550.0 & …     \\
1917.0 & 0.22 & 1929.1 & 0.02 & 1560.3 & 0.35 & 1567.0 & 0.13  \\
1947.8 & …    & 1948.7 & 0.06 & 1600.7 & 0.14 & 1603.0 & 0.03  \\
1968.6 & 0.03 & …      & …    & 1634.7 & 0.27 & 1637.0 & 0.19  \\
       &      &        &      & 1668.3 & 0.14 & 1672.0 & 0.14  \\
       &      &        &      & 1689.5 & 0.19 & 1692.0 & 0.15  \\
       &      &        &      & 1720.0 & 0.10 & 1717.0 & 0.05  \\
       &      &        &      & 1744.0 & 0.11 & 1747.0 & 0.10  \\
       &      &        &      & 1771.3 & …    & 1776.0 & …     \\
       &      &        &      & 1786.2 & …    & 1791.0 & …     \\
       &      &        &      & 1798.0 & 0.30 & 1799.0 & 0.32  \\
       &      &        &      & 1827.4 & 0.14 & 1826.0 & 0.04  \\
       &      &        &      & 1845.5 & 0.07 & 1847.0 & 0.09 \\
\hline
\end{tabular}
\end{table}
\end{appendix}

\end{document}